\documentclass[10pt]{article}

\usepackage{amsmath}
\usepackage{amssymb}

\usepackage{graphicx}

\usepackage{cite}

\usepackage{color} 

\usepackage[labelfont=bf,labelsep=period,justification=raggedright]{caption}

\makeatletter
\renewcommand{\@biblabel}[1]{\quad#1.}
\makeatother

\date{}

\newcommand{\rnaseq}{RNA-Seq}

\begin{document}

\begin{flushleft}
{\Large
\textbf{\rnaseq{} Mapping Errors When Using Incomplete Reference Transcriptomes of Vertebrates}
}
\\
Alexis Black Pyrkosz$^{1}$, 
Hans Cheng$^{1}$, 
C. Titus Brown$^{2,\ast}$
\\
\bf{1} Avian Disease Oncology Laboratory, USDA, East Lansing, MI, USA
\\
\bf{2} Microbiology Department, Michigan State University, East Lansing, MI, USA
\\
$\ast$ E-mail: Corresponding ctb@msu.edu
\end{flushleft}

\section*{Abstract}

Whole transcriptome sequencing is increasingly being used as a functional genomics tool 
to study nonmodel organisms. However, when the reference transcriptome used to calculate 
differential expression is incomplete, significant error in the inferred expression levels can 
result.  In this study, we use simulated reads generated from real transcriptomes to determine 
the accuracy of read mapping, and measure the error resulting from using an incomplete 
transcriptome.  We show that the two primary sources of counting error are 1) alternative splice 
variants that share reads and 2) missing transcripts from the reference. Alternative splice 
variants increase the false positive rate of mapping while incomplete reference transcriptomes 
decrease the true positive rate, leading to inaccurate transcript expression levels.  Grouping 
transcripts by gene or read sharing (similar to mapping to a reference genome) significantly 
decreases false positives, but only by improving the reference transcriptome itself can the 
missing transcript problem be addressed.  We also demonstrate that employing different 
mapping software does not yield substantial increases in accuracy on simulated data.  Finally, 
we show that read lengths or insert sizes must increase past 1kb to resolve mapping ambiguity.

\section*{Introduction}

Whole transcriptome sequencing using RNA sequencing (\rnaseq{}) has grown in popularity 
for studying organisms in agriculture, evolution, and medicine \cite{Martin2011,Twine2011,Kai2011}.  
Next generation sequencing (NGS) technology has evolved to the point where the cost is within 
financial range for many labs, and the resultant data can be mined for a range of applications 
\cite{Ye2011}.  However, the computational pipelines to process \rnaseq{} data were developed 
for organisms with well-studied reference transcriptomes.
In this study, 
we use simulations to assess the accuracy of a key part of this pipeline, mapping short reads to 
a reference transcriptome, and quantify the error arising from incomplete transcriptomes and 
alternative splicing.

\rnaseq{}, deep sequencing of cDNA, can be used to generate a transcriptome that includes 
alternative splice variants, and can also be used to estimate the expression level of these transcripts \cite{Wang2008,Cloonan2008,Denoeud2008,Lister2008,Maher2009,Marioni2008,Mortazavi2008,Nagalakshmi2008,Sultan2008,Yassour2009}.  Transcripts can be quantified and analyzed for differential expression between samples.  
Current \rnaseq{} approaches use shotgun sequencing with short-read technologies such as 
Illumina, in which millions or billions of 50-150 bp reads are generated from a randomly 
fragmented cDNA library.  Because these reads are not long enough to span most complete 
transcripts, expression levels must be inferred by using a reference transcriptome.

Current approaches for converting read data into transcript expression levels consist of 
1) obtaining a reference transcriptome, 2) mapping reads to the reference, and 3) applying 
statistical estimation to obtain expression levels and compare transcript expression across 
conditions.  The first step makes use of either an existing reference transcriptome or building 
a transcriptome from the \rnaseq{} reads via reference-guided or {\em de novo} assembly 
\cite{Trapnell2010,Schulz2012,Grabherr2011}.  Because many publicly available reference 
transcriptomes are incomplete \cite{Ye2011}, researchers must often assemble their own 
reference \cite{Oshlack2010}, a nontrivial task that requires building contiguous regions 
from many overlapping short, low quality reads \cite{Martin2011}.  This is usually accomplished 
using de Bruijn graphs that model reads as overlapping subsequences (k-mers) and build 
a graph of nodes that can be traversed to generate transcripts 
\cite{SurgetGroba2010,DeBruijn1946,Pevzner1989,Zerbino2008}.  Reads derived from 
low expression transcripts can be so low in abundance that they are difficult to distinguish 
from errors and therefore can be difficult to assemble \cite{Robertson2010,Zerbino2010}.  
Transcriptome assembly is particularly complicated in plants and animals because alternative 
splice variants often share exons, and distinct transcripts can be difficult to resolve 
\cite{Lander2001,Modrek2001,Johnson2003}.

The second step requires matching each read to a transcript, a process usually performed 
by a fast indexing algorithm.  Mapping performed against a reference transcriptome typically 
results in high matching accuracy between the reads and reference as reads originating 
from unassembled transcripts will be discarded or mismapped \cite{Costa2010,Stein2011}.  
Most mapping software was benchmarked with experimental genomic 
data from \emph{Homo sapiens} 
with emphasis on sensitivity as opposed to accuracy.
One critical mapping parameter is how to handle multimap reads, i.e., reads that map 
equally well to multiple locations in the transcriptome \cite{Oshlack2010}.  These are usually  caused by alternative splice variants that share exons or repetitive regions or 
recently divergent homologs \cite{Garber2011}.  Some studies have placed the number of multimap reads 
for a mammalian genome between 10 and 40\% \cite{Costa2010,Cloonan2008,Mortazavi2008}.  
The ÒdefaultÓ setting for most mapping software is to randomly select a position from the 
possible matches, whether this is the highest scoring position ÒbestÓ or the first position that 
matches a threshold score.  Alternatively, this parameter can be set to map only ÒuniqueÓ 
reads, which discards a large number of reads.  The parameter can also be set to Òmultimap,Ó 
which reports all matching positions per read to be screened by another technique.  Studies 
to statistically distribute multimap reads have had varying levels of success, starting with 
the rescue method of using unique reads to initially estimate expression 
levels, fractionally allocating multimap reads, and reestimating, and expanding to fitting 
multimap read allocations to various statistical models 
\cite{Mortazavi2008,Marioni2008,Cloonan2009,Hashimoto2009,Jiang2009,Li2010,Trapnell2010,Li2011,Trapnell2012} 
and using exon expression levels \cite{Kim2011,Richard2010}.  Some studies have 
indicated that accurately allocating all multimap reads to transcripts cannot be done because 
some splice variants are linear combinations of other splice variants \cite{Lacroix2008,Xia2011}.

The third step determines transcript/gene expression levels and finds differentially 
expressed transcripts/genes.  Because we cannot directly measure whole-transcript 
expression with current technology, a variety of sophisticated models have been used 
to infer transcript expression from mapped reads (reviewed in \cite{Trapnell2012}).

A key question for researchers working with \rnaseq{} is the extent to which expression 
analysis depends on the quality and completeness of their reference transcriptome.  An 
incomplete reference transcriptome can result in inaccurate read assignment due to missing 
transcripts, and this can bias downstream expression analyses.  A related question is 
whether longer reads or (for paired-end sequencing) larger fragment sizes can yield 
improved mapping specificity.

In this study we determine the mapping and transcript expression accuracy that can be 
expected for a \rnaseq{} experiment on a vertebrate with an incomplete reference 
transcriptome.  We first simulate small data sets for real and artificial transcriptomes 
using both single and paired-end reads and quantify the mapping error resulting from 
alternative splice variants.  We then use larger simulated read sets for real transcriptomes 
and demonstrate the effects of alternative splice variants and incomplete reference 
transcriptome on read mapping accuracy.  We next determine the error in transcript expression 
and show that grouping transcripts by gene or shared sequence can improve accuracy.  
Finally, we simulate reads of increasing length to determine what read length is needed 
for higher accuracy.

\section*{Results}

\subsection*{Read Simulations}

To evaluate the accuracy of mapping, read sets were simulated where each read was annotated 
with its originating transcript.  The transcriptomes used were either chicken, 
mouse, or randomly generated (see Methods).  
To simulate Illumina reads, single end reads were 100 bp and 
paired end reads were 100 bp with 250 bp insert size.  An error rate of 1\% was simulated by 
applying a random substitution error to the simulated reads.  Reads were generated from 
randomly selected 100 or 250 bp spans along each transcript.  Expression levels for most 
datasets were based on two models: 1) 20x average coverage with 100\% of the transcripts in 
the reference expressed, or 2) 0, 10, 100, or 1000x average coverage per transcript with 50\% 
of the transcripts in the reference randomly selected for uniformly chosen nonzero expression.

Mapping of the reads to the reference transcriptome was performed using Bowtie except where 
specified \cite{Langmead2009}.  The effects of an incomplete reference transcriptome were 
explored by incrementally removing 10\% of the transcripts from the reference transcriptome 
(randomly selected).

\subsection*{Errors from isoforms in a small random vs real transcriptome test set}

\begin{figure}[!ht]
\begin{center}
\includegraphics[width=6in]{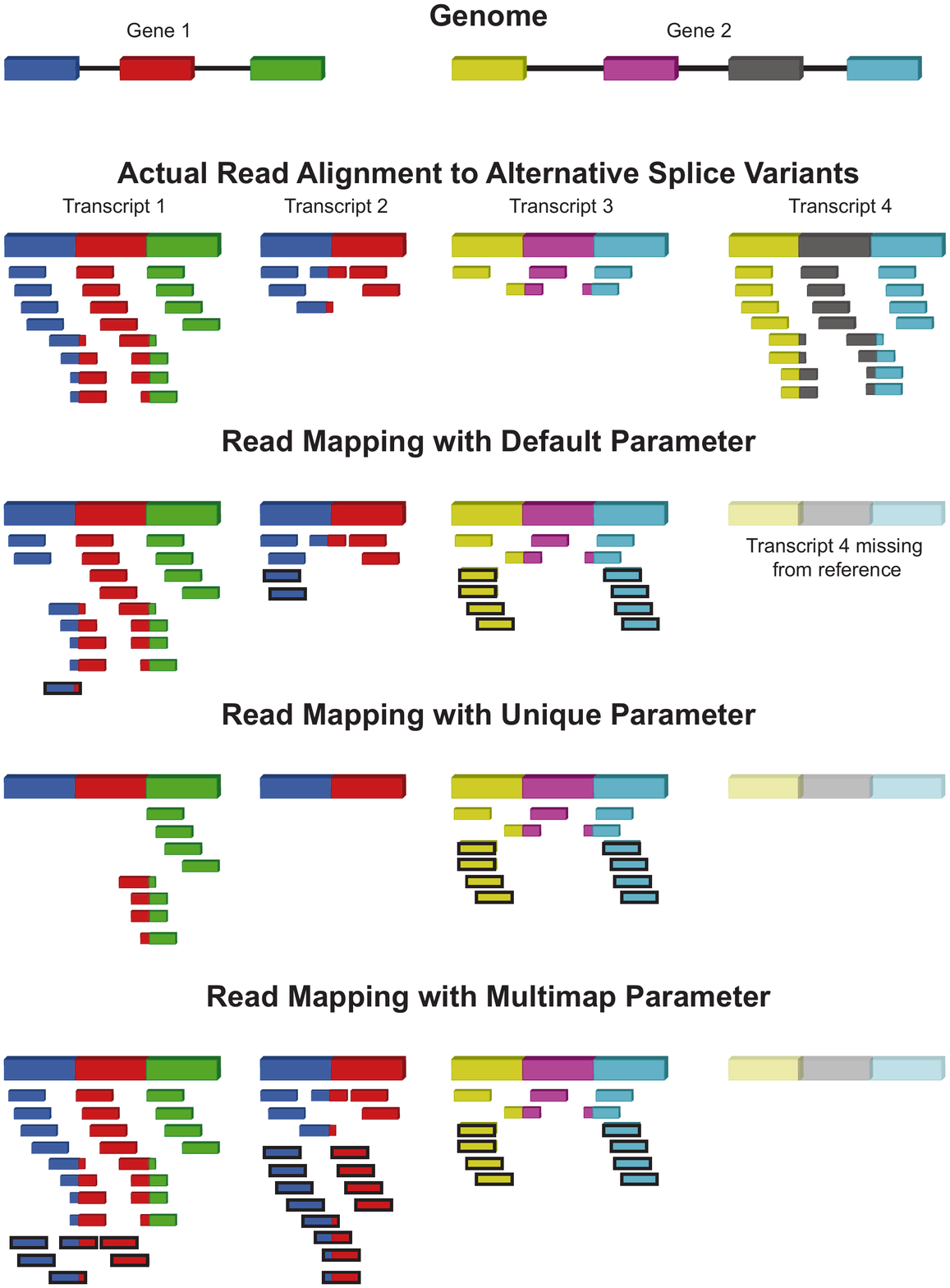}
\end{center}
\caption{
{\bf Read mapping errors when using different parameters.  Large colored rectangles indicate different exons; transparent transcripts are missing from the reference; and black outlines indicate reads that are erroneously mapped.  In the Actual Read Alignment, reads are shown mapped to their 
correct transcript.  In Read Mapping with Default Parameter, the same reads are shown but the multimap reads can be mapped to the wrong 
transcript.  Multimap reads originating from Transcript 4 are all mapped to Transcript 3 when Transcript 4 is missing from the reference transcriptome.  
In Read Mapping with Unique Parameter, the multimap reads for Transcripts 1 and 2 are removed from the analysis, thereby preventing erroneous 
mapping, but the multimap reads for Transcripts 3 and 4 are not removed because they are not known to be multimapping when Transcript 4 
missing from the reference. 
}}
\label{fig:masterMap}
\end{figure}

To characterize mapping errors and mapping parameters effects, we first generated small read 
sets from 5000 transcripts with 20x average coverage, and mapped them to a complete reference 
transcriptome (in triplicate with results averaged; variance not shown).  As shown in Table 
\ref{tab:toymodel}, mapping reads to a randomly generated transcriptome is 92\% (single end 
or SE) or 85\% (paired end or PE) correct regardless of multimap parameters, with 8\% (SE) or 
15\% (PE) of reads lost due to reads having two or more sequencing errors.  The lack of false 
positives is expected because random transcriptomes are unlikely to contain transcripts that 
share reads. When mapping reads to a randomly selected subset of the mouse transcriptome 
containing alternative splice variants, multimap reads are mismapped using either default or 
multimap parameters; false positives are prevented by using the unique parameter to eliminate 
these reads from the data set (see Figure \ref{fig:masterMap}).  Simulations of the random 
transcriptome with increasing numbers of alternative splice variants confirm that more multimap 
reads lead to higher false positive rates (see Table \ref{tab:toymodelsims}).  Mapping specificity 
is higher for paired ends, as seen with the lower false positive rate for the mouse transcriptome 
using both the default and multimap parameters.  The 8\% or 15\% unmappable reads due to 
substitution error is expected to remain constant for read sets of all sizes (due to sequencing 
platform used); but for larger reference transcriptomes containing more alternative splice variants, 
the false positive rate due to multimap reads is expected to be compounded.

\subsection*{Comparison of incomplete reference mapping results between model and nonmodel organisms}

\begin{figure}[!ht]
\begin{center}
\includegraphics[width=6in]{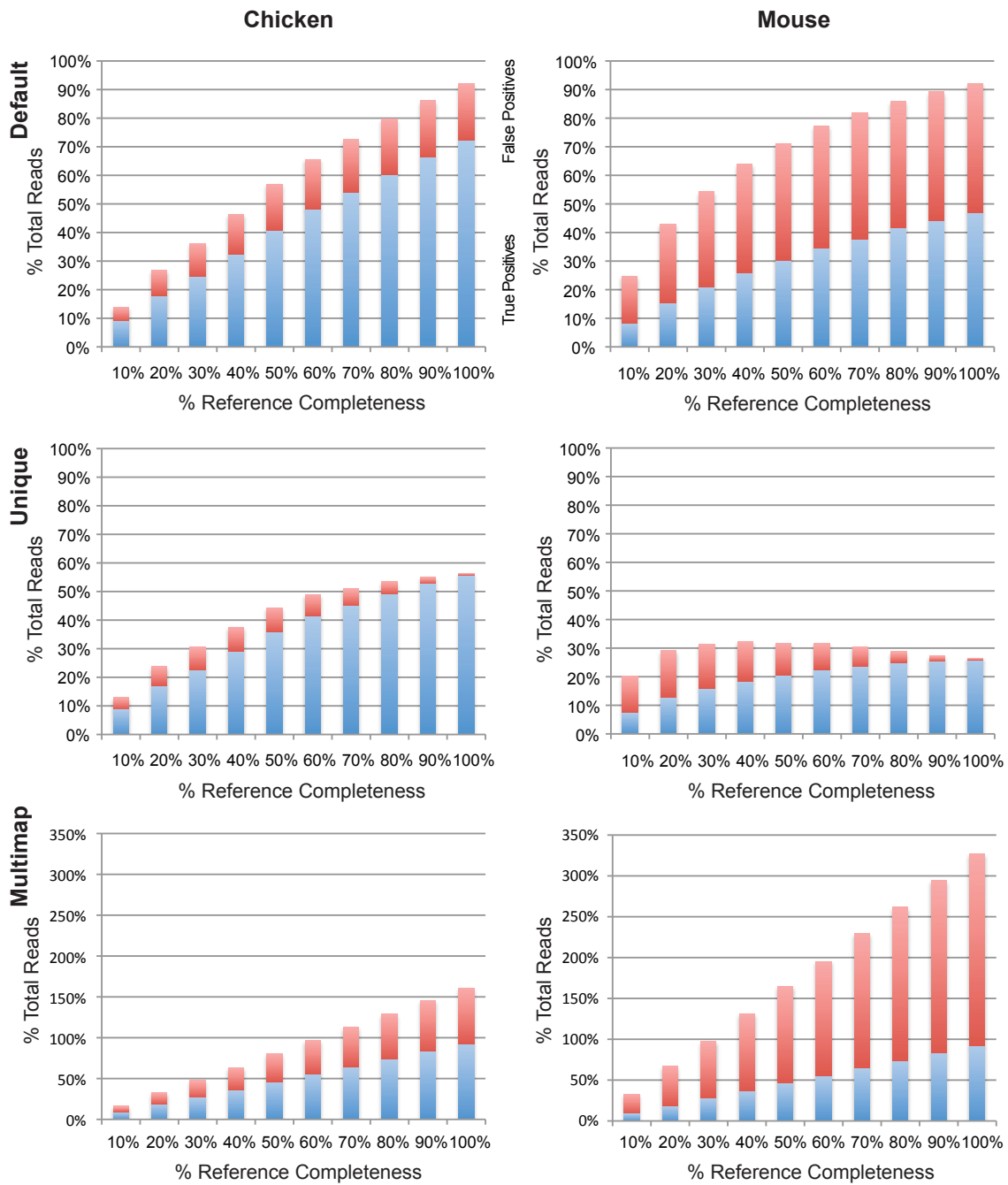}
\end{center}
\caption{
{\bf Read mapping errors for single end reads from real transcriptomes with 10\% increments of the reference transcriptome eliminated.  
Blue bars indicate true positives and red bars are false positives.  Note difference in y-axis for third row due to high false positive 
when multimapping (reads with many equivalent mapping locations).
}}
\label{fig:incRefComposite}
\end{figure}

To characterize the effects of an incomplete reference transcriptome on mapping accuracy, 
we next generated simulated read sets for \emph{Mus musculus}, which has a well-studied transcriptome
(23,153 genes and 90,956 transcripts in release 68 as of November 2012\cite{Flicek2012}), 
and \emph{Gallus gallus}, for which the reference genome is predicted to be 95\% complete 
and transcriptome is 
incomplete (16,736 genes and 23.392 transcripts in 2.1 release as of November 2012) 
with sequence missing for several microchromosomes\cite{Ye2011}. 50\% of the 
transcripts were randomly selected for expression across several orders of magnitude (10, 100, 1000) 
to model real \rnaseq{} expression levels (triplicate runs with results averaged).  
As seen in Figure 2 (top left), when 10\% increments of 
the chicken reference transcriptome are missing, the true positive rate decreases by approximately 
6-8\%, while the false positive rate remains relatively constant until the reference is more than 50\% 
incomplete.  This is due to reads being mismapped to alternative splice variants (as demonstrated 
by Transcripts 3 and 4 in Figure \ref{fig:masterMap}) when the reference is mostly complete and due 
to reads being unmappable when the reference is mostly incomplete.  This is confirmed when only 
unique reads are mapped (Figure \ref{fig:incRefComposite}, middle left); the number of false positives 
grows as the reference becomes increasingly incomplete and decreases when the reference is  
less than 30\% complete (few splice variants left in the reference).  This trend is amplified in the 
mouse data set (Figure \ref{fig:incRefComposite}, right column) by the significantly larger number 
of known alternative splice variants, leading to less than 50\% of the reads being accurately 
mapped with default parameters and less than 25\% of the data being mappable using unique 
parameters.  Because the chicken transcriptome is less complete than mouse, it is probable that 
real chicken \rnaseq{} data would more closely resemble the high false positive rates of the mouse 
data.  Paired read data sets generated similarly to the single end sets discussed above have similar 
mapping trends and accuracy as shown in Figure \ref{fig:incRefComposite} (not shown).

These errors are intrinsic to the data and current mapping algorithms.  As seen in Table
\ref{tab:mapperComp} when using three different popular Burrows-Wheeler mapping programs 
for Illumina data on sets of simulated chicken read data\cite{Langmead2009,Li2009bwa,Li2009soap2}, 
the results are nearly identical.  These results confirm that alternative splice variants cause 
high/nearly constant false positive rates while incomplete reference transcriptomes decrease the 
true positive rate.

\subsection*{Errors in transcript expression levels}

\begin{figure}[!ht]
\begin{center}
\includegraphics[width=6in]{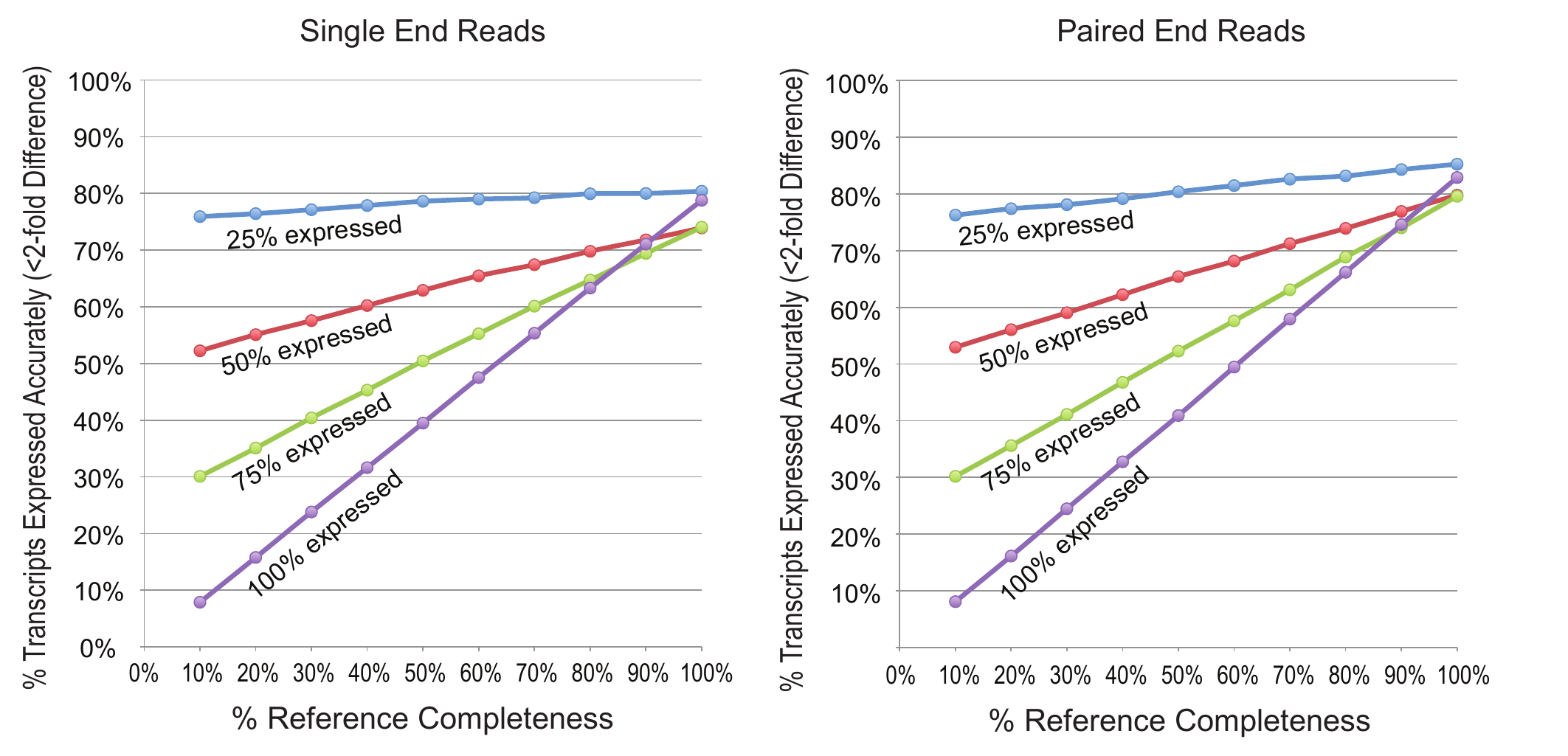}
\end{center}
\caption{
{\bf Transcript level errors for chicken transcriptome with differing levels of transcriptome expression and increasingly incomplete reference.  
Percentages indicate the amount of the transcriptome expressed (i.e. 25\% expressed indicates that 25\% of the transcripts in the chicken 
transcriptome have nonzero expression).    
}}
\label{fig:transcriptExp}
\end{figure}

\begin{figure}[!ht]
\begin{center}
\includegraphics[width=6in]{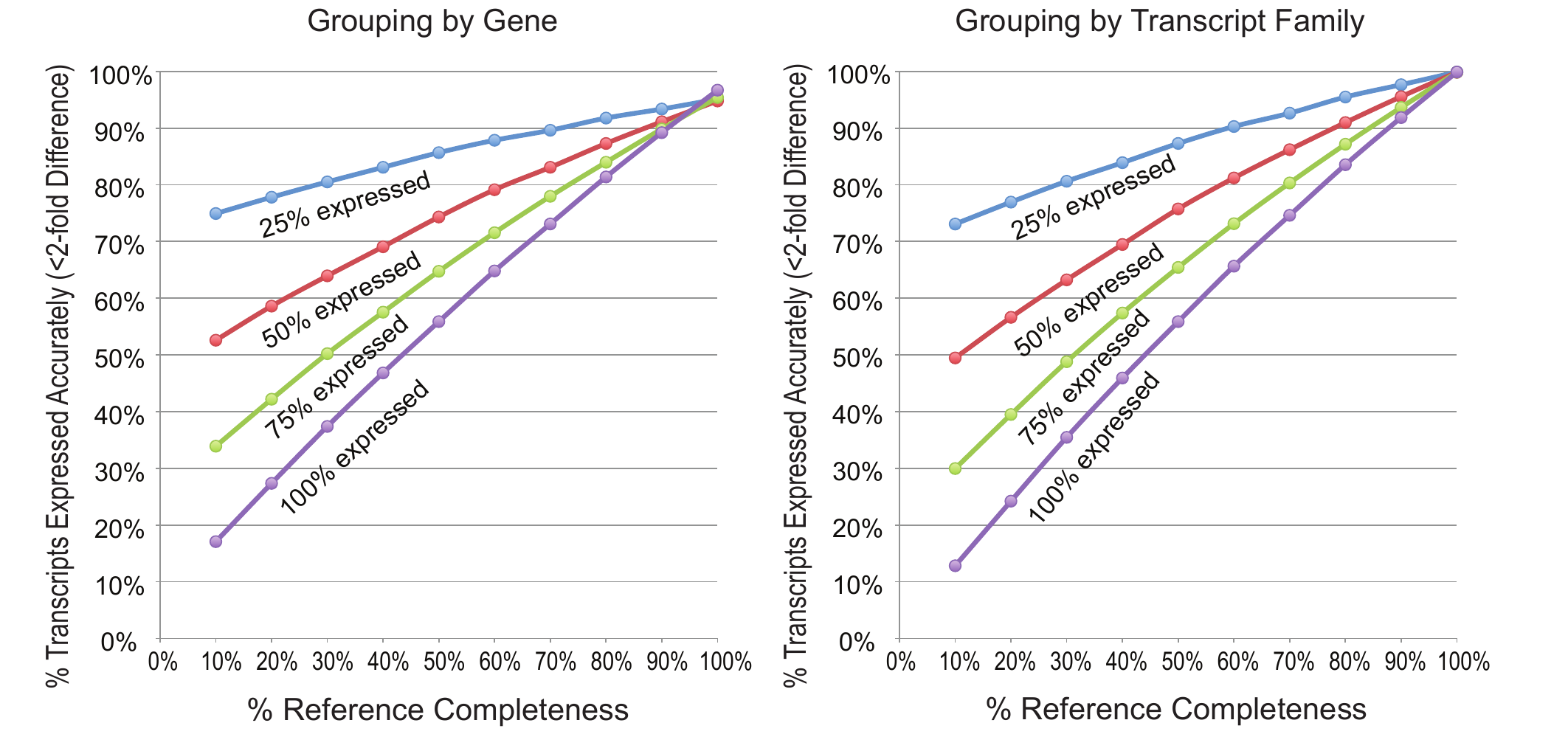}
\end{center}
\caption{
{\bf Gene and transcript family level errors for chicken transcriptome with differing levels of transcriptome expression and increasingly incomplete reference transcriptome (single end reads).
}}
\label{fig:geneExp}
\end{figure}

To determine the effect of alternative splice variants and incomplete reference transcriptomes on 
transcript expression levels, we compared the number of reads mapped to each transcript to the 
number of reads generated from each transcript in the chicken transcriptome.  As seen in Figure 
\ref{fig:transcriptExp}, regardless of the percent of the transcripts in the reference expressed, between 
15 and 27\% of the transcripts have an inaccurate expression level when mapping with a complete 
reference transcriptome.  Paired ends yield slightly greater mapping specificity and therefore slightly 
more accurate transcript expression levels.  As the reference transcriptome completeness decreases, 
the read mapping errors lead to more transcripts with inaccurate expression levels reported, with higher 
percentages of transcriptome expression leading to greater numbers of errors.  Errors from alternative 
splice variants can be eliminated by grouping transcripts either by gene (similar to genome mapping) 
if gene to transcript information is available, or grouping based on transcripts sharing multimap reads 
(see Figure \ref{fig:geneExp}).  Errors from incomplete reference transcriptomes can only be eliminated 
by improving reference transcriptomes.

\subsection*{Longer read lengths increase read uniqueness and decrease error}

\begin{figure}[!ht]
\begin{center}
\includegraphics[width=6in]{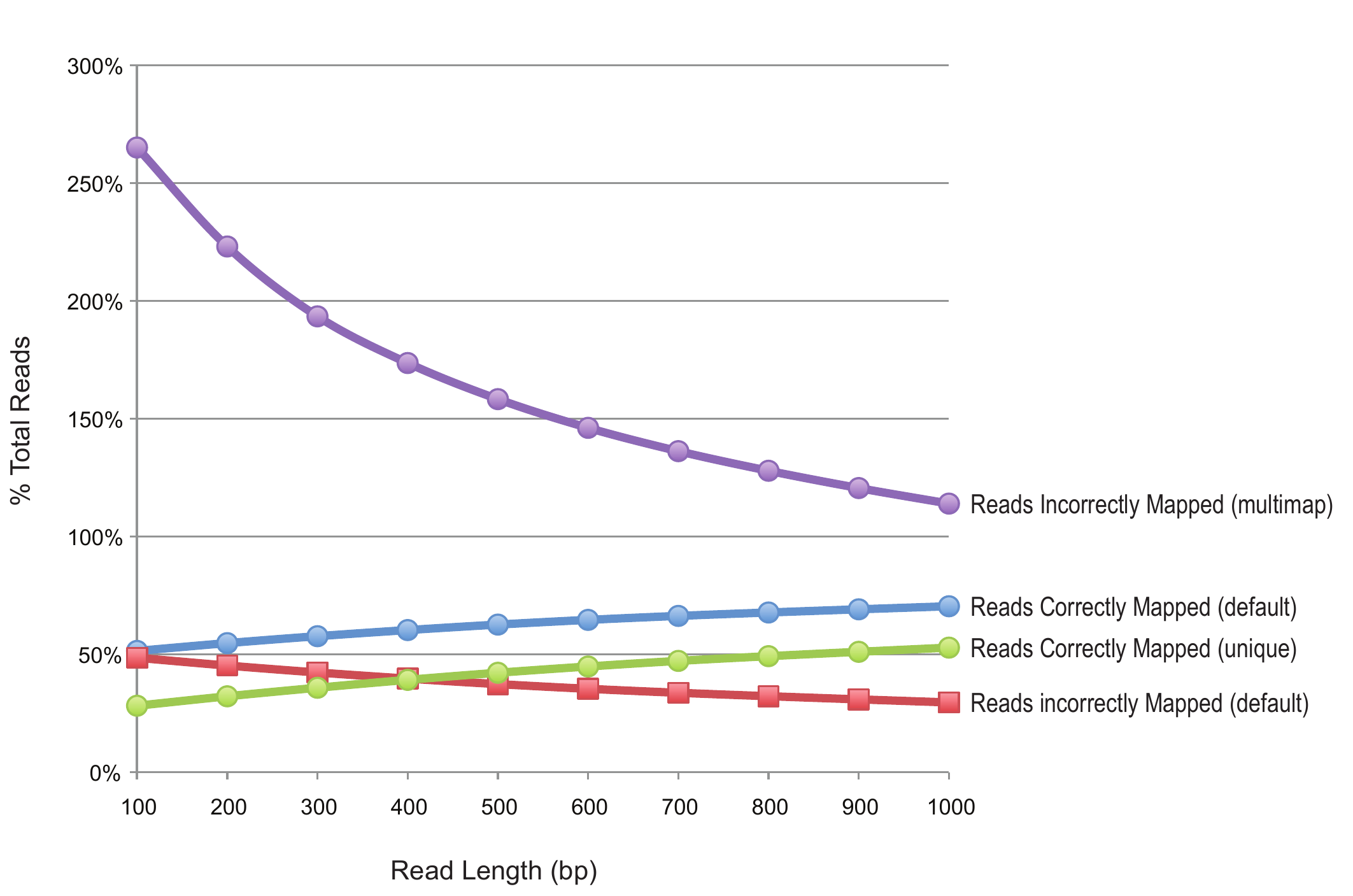}
\end{center}
\caption{
{\bf Read mapping errors for single end reads from 20x coverage mouse transcriptome (no substitution errors) with increasing read length.
}}
\label{fig:longSeReads}
\end{figure}

\begin{figure}[!ht]
\begin{center}
\includegraphics[width=6in]{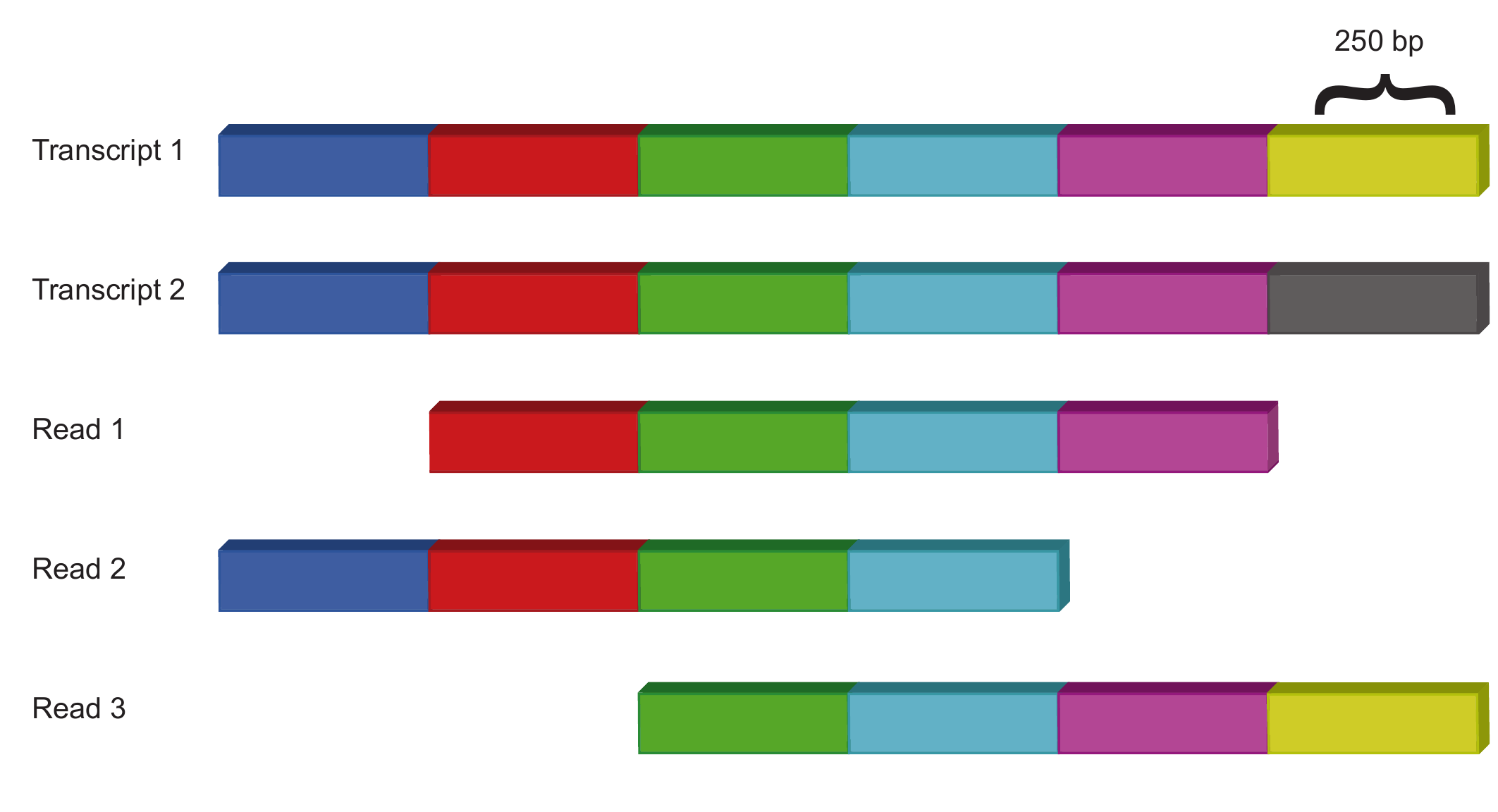}
\end{center}
\caption{
{\bf Hypothetical example of 1 kb multimap reads.  Only Read 3 can be uniquely mapped due to the unique exon in Transcript 1.
}}
\label{fig:longSeReadsExample}
\end{figure}

\begin{figure}[!ht]
\begin{center}
\includegraphics[width=6in]{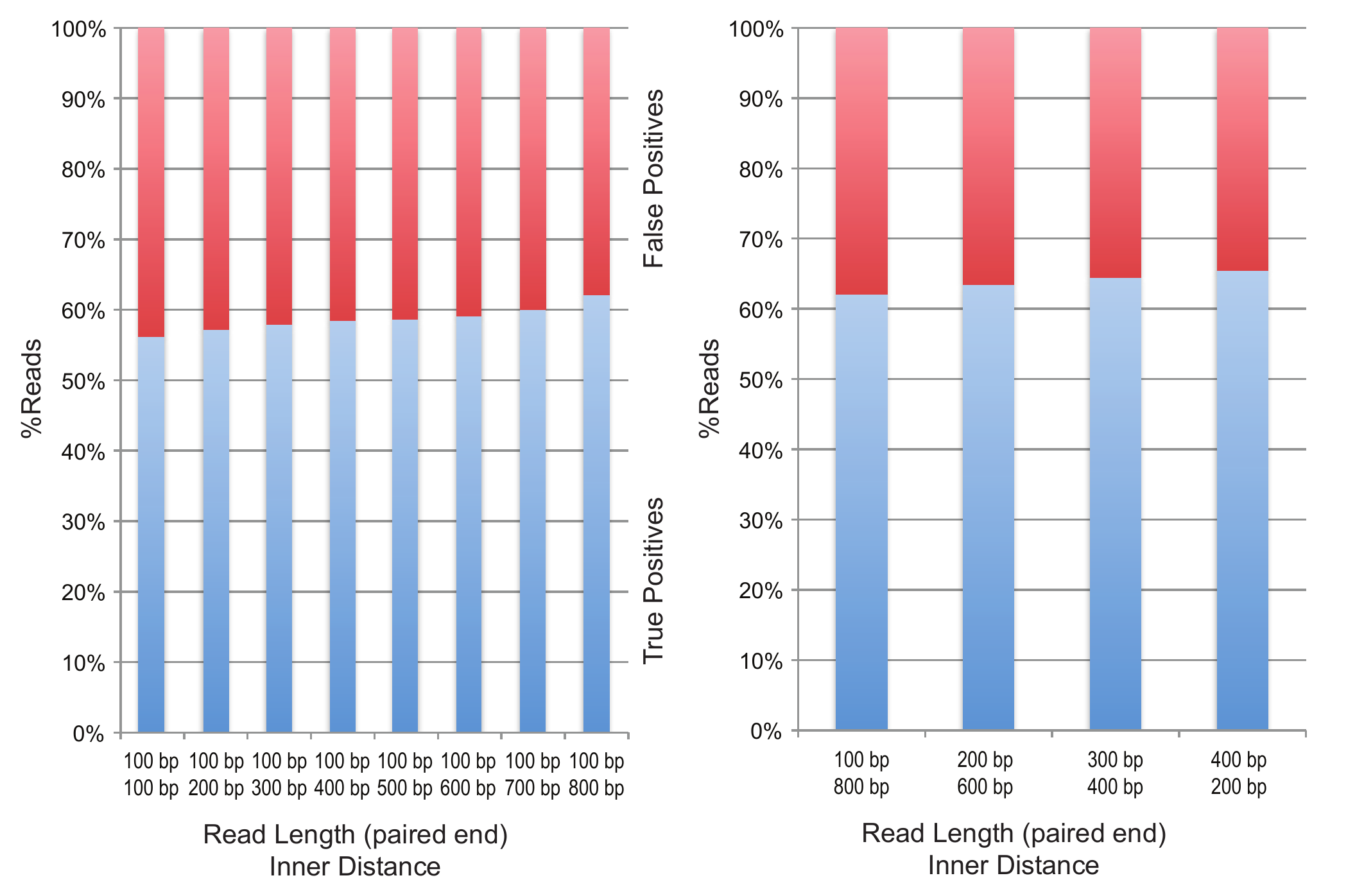}
\end{center}
\caption{
{\bf Comparison of read mapping errors for paired end reads from 20x coverage mouse transcriptome (no substitution error) with default parameter.  
Left, read lengths remain 100 bp for increasing fragment lengths.  Right, read lengths increase for 1 kb fragments.
}}
\label{fig:longPeReads}
\end{figure}

The most commonly cited solution to the current high mapping error problem (and incomplete assembly 
problem) is to develop new sequencing technologies that generate longer reads.  Illumina currently has 
single-end read lengths between 100 and 250 bp in length, Roche 454 can achieve 400-800 bp, and Pacific Biosciences is 
developing a method that will yield 10000 bp, albeit with lower numbers of reads.  To determine if these lengths are sufficient to increase 
the uniqueness of reads, we simulated single and paired ends reads at various lengths and determined 
the read mapping errors.  As seen in Figure \ref{fig:longSeReads}, 20\% more reads are unique when read 
size is increased from 100 bp to 1 kb.  However, even at 1 kb, the longest read that Bowtie can currently 
handle, reads map with only 70\% accuracy, indicating that single end reads will need to lengthen 
more than 10-fold to solve the mapping inaccuracy problem (see Figure \ref{fig:longSeReadsExample}).  
The mean length of transcripts in both the mouse and chicken transcriptomes is roughly 2 kb with a 
sizeable standard deviation; 1 kb reads are insufficient to sequence most whole transcripts and therefore 
will not necessarily be unique to long alternative splice variants.  While paired ends from fragments greater 
than 1 kb in length could achieve read uniqueness by spanning longer stretches of sequence, Figure
 \ref{fig:longPeReads} (left) shows that in comparison to single end reads of length 1 kb, paired end reads 
 of 100 bp have significantly lower uniqueness, gaining only about 1\% accuracy for every 100 bp added 
 to the inner distance.  As seen in Figure \ref{fig:longPeReads} (right), a combination of increasing 
 fragment length and increasing read length can achieve accuracy rates only slightly lower than those 
 of long single end reads; 400 bp paired end reads with a 200 bp inner distance are only 5\% less 
 accurate than 1 kb single end reads.

\section*{Discussion}

As \rnaseq{} becomes less expensive and more computational tools are developed, the focus on 
analysis should shift to which questions can be addressed with reasonable accuracy.  \rnaseq{} reads 
mapped to a complete reference transcriptome can be used to gather gene expression data, but 
transcript expression analyses will be skewed by erroneous mapping from alternative splice variants.  
For eukaryotes whose transcriptomes contain many splice variants, incomplete or erroneous 
transcriptomes can cause significant mismapping of reads and resulting misestimation of expression 
levels.

\subsection*{Read mapping accuracy is mode dependent and dependent on transcriptome completeness}

The dominant effect on true and false positive rates in mapping is from the mapping parameter 
used: multimap read mapping, in which all possible mappings are reported, has the highest true 
positive rate, but also comes with substantial false positive matches.  The true positive rates of both 
unique-only and default/random-assignment mapping modes are considerably lower than for 
multimap modes, although the false positive rates are also significantly lower.  Thus the choice of 
mode represents a tradeoff between sensitivity and specificity in read mapping.

The completeness of the reference transcriptome interacts significantly with the mapping mode.  
The true positive rate for multimap mode is dependent solely on the correct transcript being in the 
reference, while the default and unique modes have substantially lower true positive rates for 
incomplete transcriptomes, because both modes depend on choosing or rejecting mappings from 
among the available transcripts. In contrast, the false positive rate for multimap and default modes 
is high for transcriptomes with many splice variants, while the unique mapping mode false positive 
rate decreases substantially as more of the transcriptome is available.

\subsection*{Calculating transcript-specific expression levels directly from \rnaseq{} is challenging}

Under a simple model of transcript expression in which expression levels are chosen randomly 
and uniformly from four expression levels, gene- or transcript-family-level expression calculations 
approach 100\% accuracy when a nearly complete transcriptome is used (Figure \ref{fig:geneExp}).  
However, even with a nearly complete chicken reference transcriptome, only about 80\% accuracy 
can be achieved for single transcript measurements.  Moreover, the more transcripts that are expressed, 
the less accurate the expression calculations are, especially in tandem with a less-than complete 
chicken transcriptome (Figure \ref{fig:transcriptExp}).   It is important to note that the chicken transcriptome 
being used here is likely lacking many real splice variants compared to mouse or human, which are 
comparatively better studied; the expression accuracy for single transcripts will inevitably decrease 
with more splice variants to confuse read mapping.

\subsection*{Longer reads and fragment lengths do not result in perfect mapping}

Somewhat to our surprise, neither 1 kb reads nor paired end reads from 1 kb fragments resolve the 
mapping problem.  Even with 1 kb reads and a complete transcriptome, multimap mapping has about 
a 20\% false positive rate for mouse (Figure \ref{fig:longSeReads}), while neither default nor unique 
mapping mode can achieve better than 70\% true positives.  This problem emerges directly from the 
length of many transcripts and the distance between exons; e.g. see Figure \ref{fig:longSeReadsExample}.

Even when longer reads are available (as from the Pacific Biosciences SMRT technology), sampling 
depth is also critically important Ð rare transcripts cannot be observed systematically with low 
coverage. One study on B-cells suggests that between 100 and 500 million observations are 
necessary for accurate transcript quantification \cite{Toung2011}.  Current sequencing technology 
is not yet close to providing both long reads and high sampling.

\subsection*{Limits to mapping accuracy also imply limits to \rnaseq{} assembly accuracy}

In the absence of deep sampling with long reads, most transcriptomes are inferred from \rnaseq{} 
using {\em de novo} or reference-based assembly methods.  Our read mapping simulations suggest that 
there are fundamental limits to the ability of assembly methods to determine complete transcripts 
from \rnaseq{}, and indeed all assembly methods use more or less explicit models to deconvolve 
reads and read counts into distinct isoforms \cite{Trapnell2010,Schulz2012,Grabherr2011}.  Absent 
direct observations, these inferences are necessarily of unknown quality; cross validation would 
require the direct observation of long-range exon-exon correlations in single molecules, and none 
of \rnaseq{}, microarrays, or qPCR can provide such observations.  In practice, most methods simply 
seek to maximize the concordance between the observed information and the assembled transcripts, 
which is almost certainly sufficient to reconstruct the majority of exons and exon-exon junctions 
correctly.  This is confirmed by the strong correlation between results from \rnaseq{}, ESTs, and 
full-length cDNA sequencing in many model organisms, as well as bioinformatic validation of 
protein-coding sequences.  Thus we believe that it is unlikely that many transcriptomes are systematically 
inaccurate on a large scale.

The biggest uncertainty in reference transcriptomes is likely to be due to the presence of 
low-expressed isoforms,  many of which may come from ÒnoisyÓ splicing and are biologically 
irrelevant \cite{Pickrell2010}.  Unfortunately accurate measurement of the expression levels of 
rare transcripts across conditions is dependent on accurate reference transcriptomes, leading to 
a chicken-and-egg problem in our ability to evaluate biological relevance.  We know of no quantification 
methods that report measurement uncertainty without relying on a preexisting reference.

Separately, the confounding effect of too complete a reference on correct mapping and expression 
inference may be significant.  If many unexpressed splice variants are present in the reference 
transcriptome used for quantification, they may result in inaccurate expression calculations.  A variety 
of adaptive shrinkage approaches have been applied to this problem 
\cite{Li2011,Li2011isolasso,Nguyen2012} but they are again reference dependent.

\section*{Conclusion}

Our basic conclusion is that \rnaseq{}-based expression analysis is strongly dependent on 
the quality and completeness of the reference transcriptome, suggesting that new \rnaseq{} data 
sets be integrated into existing references prior to expression analysis.  While many tools exist for 
this purpose, it is difficult to evaluate their performance in light of the limits we observed for correct 
transcript mapping and their implications for correct transcript reconstruction.

Another important observation, already made by several others, is that differential expression can 
be most accurately calculated at a gene- or transcript-family-level, while isoform-level expression 
is subject to significant uncertainty \cite{Trapnell2010,Trapnell2012}.  Moreover, this is unlikely to change until technologies that permit 
deep sampling with 2 kb reads are readily available.

\section*{Materials and Methods}

\subsection*{Random transcriptome and alternative splice variant simulations}

Python scripts were used to generate transcripts with randomly selected A, C, G, and T characters 
with a minimum length of 100 and maximum of 5000 bp.  Alternative splice variants were simulated 
by randomly selecting existing transcripts from a transcriptome and performing one of five operations: 
truncation (removal of a random length of characters from the end) or skipping an exon at the 3' end, 
extension (addition of a random length of characters to the end) or adding an exon at the 3' end, 
insertion (addition of a random length of characters at a randomly selected position in the middle) or 
splicing in a new exon or failure to remove an intron, deletion (removal of a random length of characters 
at a randomly selected in the middle) or exon skipping, or substitution (replacement of a random length 
of characters at a randomly selected position with a randomly generated length of characters) or swapping 
exons.  No limit was placed on the number of times a transcript could be used to form a variant.  
Variants generated early in the process could be used to form further variants later.

\subsection*{Read generation}

Python scripts were used to generate single and paired end reads from reference transcriptomes, 
either random transcriptomes generated as above or those downloaded from Ensembl in March 
2012 \cite{Flicek2012}.  Reads were generated by randomly selecting start positions in a given 
transcript that were at least one read length from the end of the transcript (or two read lengths + 
inner distance) and copying the sequence up to given read length.  Read lengths were 100 bp both 
for single and paired end reads except where mentioned (see Results).  A 1\% random substitution 
error was added by randomly selecting 1\% of the positions in the read and substituting the correct 
nucleotide with one of the other three nucleotides.  The number of reads generated per transcript to 
achieve the given average coverage was determined by the equation: number of reads = (transcript 
length) / (read length) * (coverage).  Average coverage was either 20x for all transcripts in the 
transcriptome or were randomly selected from the set {0, 10, 100, 1000} according to the percentage 
of the reference expressed (the probabilities associated with the nonzero elements were equal).  Scripts 
for read generation and random transcriptome generation with splice variants are available at github 
(https://github.com/ablackpz/Simulate-mRNASeq-Reads).

\subsection*{Mapping and accuracy}

Mapping was performed using Bowtie except where specified.  The default parameters were selected 
according to the bowtie tutorial, the unique parameter was invoked as ``-m 1'', and the multimap parameter 
was ``-a''.  Paired ends were mapped with insert size between 0 and 1000.  Mapping accuracy was assessed 
for each read by comparing the transcript identifier from which each read originated with the transcript 
identifier reported for that read by the mapping program.  A match was classified as a true positive and 
mismatch as a false positive.  True/false negatives were not addressed in this study.  Transcript expression 
accuracy was assessed by counting the number of reads generated from a given transcript (actual) and 
the number of reads reported for that transcript from the mapping program (experimental).  Values where actual 
and experimental values differed more than 2-fold were considered to be erroneous.  Gene 
expression accuracy was determined by using transcript/gene relationships as reported by Ensembl 
to group transcripts by gene.  

Similar to the exon union method \cite{Garber2011}, the number of reads generated for transcripts 
within a given transcript group by gene and number of reads mapped transcripts within that 
group were counted and erroneous expression assessed similarly to above.  Transcript family 
expression accuracy was determined by first determining the transcript grouping by multimapping 
reads relationships.  Overlapping reads were systematically generated for each transcript with the 
resulting read set containing all possible reads for the reference (exact match).  This large read set 
was then multimapped to the reference transcriptome to determine which reads could map to 
multiple transcripts.  Transcripts sharing reads were grouped together.  Reads per transcript 
group were counted for both initial read generation and mapping and erroneous expression 
was determined as above.

\section*{Acknowledgments}

\bibliography{refs}

\section*{Figure Legends}


\section*{Tables}

\begin{table}[!ht]
\caption{
\bf{Read mapping errors for single (SE) and paired end (PE) reads from random (simulated) and real transcriptomes }}
\begin{tabular}{|c|c|c||c|c||c|c||c|c||c|}
\hline
Organism & Num Trans & Error  & TP (d) & FP (d) & TP (u) & FP (u) & TP (m) & FP (m)\\
\hline
Random (SE) & 5000 & 1\% & 92\%& 0\%&92\%&0\%&92\%&0\%\\
\hline
Mouse (SE) & 5000 & 1\% & 87\%& 5\%&81\%&0\%&92\%&12\%\\
\hline
Random (PE) & 5000 & 1\% & 92\%& 0\%&92\%&0\%&92\%&0\%\\
\hline
Mouse (PE) & 5000 & 1\% & 87\%& 5\%&81\%&0\%&92\%&12\%\\
\hline
\end{tabular}
\begin{flushleft} Mapping parameters are default (d), unique (u), and multimap (m).  True positives are reads 
that were successfully mapped to their originating transcript.  False positives are reads that were 
mapped to other transcripts (even if the read was an exact match to the alternate transcript).  
\end{flushleft}
\label{tab:toymodel}
\end{table}

\begin{table}[!ht]
\caption{
\bf{Read mapping errors for single end reads from random (simulated) transcriptomes with alternative splice variants artificially generated}}
\begin{tabular}{|c|c|c||c|c||c|c||c|c||c|}
\hline
Organism & Num Trans & Error  & TP (d) & FP (d) & TP (u) & FP (u) & TP (m) & FP (m)\\
\hline
0\% Isoforms & 5000 & 1\% & 92\% & 0\% & 92\% & 0\% & 92\% & 0\%\\
\hline
10\% Isoforms & 5000 & 1\% & 86\% & 6\% & 81\% & 0\% & 92\% & 14\%\\
\hline
20\% Isoforms & 5000 & 1\% & 80\% & 12\% & 70\% & 0\% & 92\% & 33\%\\
\hline
30\% Isoforms & 5000 & 1\% & 74\% & 18\% & 62\% & 0\% & 92\% & 60\%\\
\hline
40\% Isoforms & 5000 & 1\% & 67\% & 25\% & 53\% & 0\% & 92\% & 105\%\\
\hline
50\% Isoforms & 5000 & 1\% & 60\% & 32\% & 45\% & 0\% & 92\% & 217\%\\
\hline
60\% Isoforms & 5000 & 1\% & 55\% & 37\% & 39\% & 0\% & 92\% & 303\%\\
\hline
70\% Isoforms & 5000 & 1\% & 48\% & 45\% & 33\% & 0\% & 92\% & 732\%\\
\hline
80\% Isoforms & 5000 & 1\% & 42\% & 50\% & 27\% & 0\% & 92\% & 970\%\\
\hline
90\% Isoforms & 5000 & 1\% & 37\% & 56\% & 23\% & 0\% & 92\% & 4722\%\\
\hline
\end{tabular}
\begin{flushleft} Mapping parameters are default (d), unique (u), and multimap (m).  Percentages indicate the amount of 
the simulated transcripts that were generated from other, randomly generated transcripts. 
\end{flushleft}
\label{tab:toymodelsims}
\end{table}

\begin{table}[!ht]
\caption{
\bf{Comparison of Three Common Mapping Programs on the Same Chicken Data Sets }}
\begin{tabular}{|c|c||c|c||c|c||c|c||c|}
\hline
Organism & Num Trans & Bowtie TP (d) & FP (d) & BWA TP (d) & FP (d) & SOAP2 TP (d) & FP (d)\\
\hline
Chicken & 100\% & 78\%& 22\%&78\%&20\%&78\%&22\%\\
Chicken & 90\% & 72\%& 21\%&72\%&20\%&72\%&21\%\\
Chicken & 80\% & 65\%& 22\%&65\%&21\%&65\%&22\%\\
Chicken & 70\% & 58\%& 22\%&58\%&21\%&58\%&22\%\\
Chicken & 60\% & 51\%& 20\%&50\%&19\%&51\%&20\%\\
Chicken & 50\% & 44\%& 19\%&44\%&18\%&44\%&19\%\\
Chicken & 40\% & 36\%& 16\%&37\%&16\%&36\%&17\%\\
Chicken & 30\% & 27\%& 13\%&27\%&13\%&27\%&12\%\\
Chicken & 20\% & 19\%& 11\%&19\%&11\%&19\%&11\%\\
Chicken & 10\% & 9\%& 5\%&9\%&6\%&9\%&5\%\\
\hline
\end{tabular}
\begin{flushleft} Comparison of Bowtie, BWA, and SOAP2 mapping programs on the same simulated reads for error-free chicken read sets (triplicate and averaged) with decreasing completeness of the reference transcriptome, showing equivalent results.
\end{flushleft}
\label{tab:mapperComp}
\end{table}

\end{document}